\documentclass[aps,pre,twocolumn,showpacs]{revtex4}
\usepackage{graphics,graphicx,subfigure}
\usepackage{amsmath}

\newcommand{\ben}{\begin{enumerate}}
\newcommand{\een}{\end{enumerate}}
\newcommand{\bit}{\begin{itemize}}
\newcommand{\eit}{\end{itemize}}
\newcommand{\be}{\begin{equation}}
\newcommand{\ee}{\end{equation}}
\newcommand{\bdm}{\begin{displaymath}}
\newcommand{\edm}{\end{displaymath}}
\newcommand{\bea}{\begin{eqnarray}}
\newcommand{\eea}{\end{eqnarray}}
\newcommand{\f}[1]{\fbox}

\begin{document}


\title{Rebuttal to: Has dark energy
really been discovered in the Lab?}

\author{Christian Beck}
\affiliation{
School of Mathematical Sciences \\ Queen Mary, University of
London \\ Mile End Road, London E1 4NS, UK}
\email{c.beck@qmul.ac.uk}
\homepage{http://www.maths.qmul.ac.uk/~beck}
\author{Michael C. Mackey}
\affiliation{Centre for Nonlinear Dynamics in Physiology and
Medicine \\ Departments of Physiology, Physics and Mathematics \\
McGill University, Montreal, Quebec, Canada}
\email{michael.mackey@mcgill.ca}
\homepage{http://www.cnd.mcgill.ca/people_mackey.html}
\altaffiliation{ also: Mathematical Institute, University of
Oxford, 24-29 St Giles', Oxford OX1 3LB, UK}


\pacs{74.81.Fa; 98.80.-k; 03.70.+k}

 \vspace{2cm}

\begin{abstract}
We argue that a recent discussion of  Jetzer and Straumann [Phys.
Lett. B {\bf 606}, 77 (2005)] relating  the measured noise spectrum
in Josephson junctions to van der Waals forces is incorrect. The
measured noise spectrum in Josephson junctions is a consequence of
the fluctuation dissipation theorem and the Josephson effect and has
nothing to do with van der Waals forces. Consequently, the argument
of Jetzer and Straumann does not shed any light on whether dark
energy can or cannot be measured using superconducting Josephson
devices. We also point out that a more recent paper of Jetzer and
Straumann [Phys. Lett. B {\bf 639}, 57 (2006)] claiming that
`zeropoint energies do not not show up in any application of the
fluctuation dissipation theorem' violates the standard view on the
subject.
\end{abstract}

\maketitle

\section{Introduction}

Recently we  hypothesized that if vacuum fluctuations underly dark
energy then this effect could be detected experimentally  using
resistively shunted Josephson junctions \cite{bm}. Our suggestion
was based on an experiment by Koch et al. \cite{koch}, who have
shown that superconducting Josephson devices have  a noise
spectrum consistent with theoretical predictions
\cite{kochtheory} based on a generalized treatment of quantum
fluctuations by Callen and Welton \cite{callen}. Subsequently,
our paper was criticized by Jetzer and Straumann \cite{jetzer,
jetzer2}, who claimed there is no basis for our hypothesis.

In this note we argue that the logic behind the Jetzer and Straumann
criticism \cite{jetzer} is misleading. Their paper \cite{jetzer} is
based on an equilibrium van der Waals model that is not applicable
to our system. We also deal with a new version of their criticism
\cite{jetzer2} and show that the view expressed in \cite{jetzer2},
namely that the noise in Josephson junctions has nothing to do with
zeropoint energies, is in apparent contrast to the standard
treatments  dealing with quantum noise in Josephson junctions
\cite{landau, kogan, gardiner}.

Our conclusion is that the arguments presented by Jetzer and
Straumann do not shed any light on a possible relation between
quantum noise and dark energy. Rather, experimental tests are
necessary, which will be performed in the near future
\cite{warburton, barber}.

\section{The data and the theory}

Koch et al. \cite{kochtheory} derived the power spectrum
$S(\omega)$ (units of A$^2$/Hz) describing the measured current
noise in a resistively shunted Josephson junction in the form
    \be
    S(\omega) =\frac{4}{R} \left[
    \frac{ \hbar\omega}{2}+\frac{ \hbar\omega}{\exp (\hbar\omega/kT)-1} \right],
    \label{eq:power}
    \ee
where $R$ (ohms) is the shunt resistor and $T$ is the absolute
temperature. The experimental work of Koch et al. \cite{koch}
convincingly demonstrated  that Equation (\ref{eq:power}) fits the
experimental data  $S(\omega)$ as a function of $\omega = 2 \pi \nu$
between $\nu = 0$ and $\nu = 6 \times 10^{11}$ Hz at 1.6 and 4.2 K.

From a formal point of view, the expression in brackets of Equation
(\ref{eq:power}) is  the mean energy
\begin{equation}
    {\bar U}(\nu,T) = \frac{1}{2} h \nu  + \frac{h \nu }{\exp(h \nu
    /kT)-1},
     \label{eq:energy}
\end{equation}
of an oscillator with frequency $\nu$ at temperature $T$. For low
temperatures the spectrum $S(\omega)$ is dominated by the linear
term in $\omega$, which can be attributed to the effects of vacuum
(zero-point) fluctuations \cite{gardiner}.  As the temperature is
increased  the second term, which is identical to the ordinary
Bose-Einstein statistics, plays an ever larger role in $S(\omega)$.

\section{The hypothesis and the criticism}

If we take the customary expression for the energy per unit volume
at a frequency $\nu$ and temperature $T$
\begin{equation}
\rho(\nu,T) = \frac{8 \pi  \nu^2}{c^3} {\bar U}(\nu,T)
\end{equation}
then 
    \be
    \rho(\nu,T) = \frac{8 \pi \nu^2}{c^3}
    \left [ \frac{1}{2}h \nu  + \frac {h \nu }{\exp(h \nu /kT)-1} \right
    ]
     \label{eq:nu-spectrum}
    \ee
In Equation (\ref{eq:nu-spectrum}) the first term
\begin{equation}
\rho_{vac}(\nu)=\frac{4\pi h\nu^3}{c^3} \label{eq:vac}
\end{equation}
is due to the zeropoint fluctuations, while the second term
\begin{equation}
\rho_{rad}(\nu , T)=\frac{8\pi h \nu^3}{c^3}
\frac{1}{\exp{(h\nu/kT)}-1} \label{eq:rad}
\end{equation}
is simply the photonic black body spectrum. Equation
(\ref{eq:nu-spectrum}) suffers from the embarrassing prediction that
there should be an infinite amount of energy per unit volume, since
\begin{displaymath}
    \lim_{\nu_c \to \infty} \int_0^{\nu_c}  \rho(\nu,T) d\nu
    \nonumber
\end{displaymath}
is divergent.  Indeed, writing
\begin{equation}
\rho(\nu,T)=\rho_{vac}(\nu) +\rho_{rad}(\nu,T), \label{4}
\end{equation}
it is easily seen that the divergence is a consequence of the
temperature independent vacuum fluctuation term because
\begin{equation}
\int_0^\infty \rho_{rad}(\nu,T)d\nu =
\frac{\pi^2k^4}{15\hbar^3c^3} T^4 \label{eq:sb-law}
\end{equation}
simply yields the customary Stefan-Boltzmann law. To circumvent
this divergence, we suggested in \cite{bm} that Equation
(\ref{eq:vac}) is only valid up to a certain cutoff frequency
$\nu_c$ so that the total energy associated with $\rho_{vac}(\nu)$
is given by
\begin{equation}
     \int_0^{\nu_c}  \rho_{vac}(\nu) d\nu =
    \frac {4   \pi  h}{c^3} \int_0^{\nu_c} \nu^3
    d\nu = \frac {   \pi  h}{c^3}  \nu_c^4.
    \label{eq:total-2}
\end{equation}
We noted that  a future experiment could examine whether the
measured vacuum fluctuations in Fig.~1 of \cite{bm}
might be a signature of dark
energy. If so, one would expect to see a cutoff in the measured
spectrum at
\begin{equation}
    \nu_c \simeq (1.69 \pm 0.05) \times 10^{12} \quad \mbox{Hz},
\label{eq:cutoff}
\end{equation}
where this value of $\nu_c$  is obtained by setting
\begin{equation}
    \frac{   \pi  h}{c^3}  \nu_c^4 \simeq \rho_{dark}
=(3.9 \pm 0.4) \mbox{GeV}/\mbox{m}^3
\end{equation}
($\rho_{dark}$ is the currently observed dark energy density in the
universe \cite{bm}).

Jetzer and Straumann \cite{jetzer} have  criticized the hypothesis
of \cite{bm} based on two different points. In their own words:

\bit
\item Point 1. `` $\cdots$ the spectral density
originally comes from a simple rational expression of Boltzmann
factors, which are not related to zero-point energies."

To illustrate their point, Jetzer and Straumann consider a
simplified model of the van der Waals force between two harmonic
oscillators and calculate the response of the system to distance
changes. Their result is independent of zero-point energies of the
two oscillators and from this they conclude that the same also holds
for the measured spectrum (\ref{eq:power}) in Josephson junctions.

\item Point 2. ``
...the absolute value of the zero-point
energy of a quantum mechanical system has no physical meaning
when gravitational coupling is ignored.
All that is measurable are changes of the zero-point energy
under variations of system parameters or of external couplings, like
an applied voltage."

Based on this general statement,
Jetzer and Straumann
claim that experiments based on Josephson junctions are unable to
detect dark energy since only differences in vacuum
energy would be physically relevant.

\end{itemize}

Here we argue in Section \ref{sec:pt1} that Point 1 is misleading
since the observed spectra in Josephson junctions have nothing to do
with van der Waals forces. In Section \ref{sec:pt2} we argue that
Point 2 is theoretically unclear (since the quantum noise in
Josephson junctions has not been shown to be renormalizable) but
experimentally testable.

\section{Point 1}\label{sec:pt1}
The justification of Point 1 of  Jetzer and Straumann \cite{jetzer}
is based on an equilibrium statistical mechanical model for the van
der Waals interaction between two identical harmonic oscillators.
The authors point out that a simple transformation can decouple the
oscillators.  The ground state of the decoupled system is the sum of
the zero-point energies of the two decoupled oscillators and the
corresponding van der Waals force is independent of the zero-point
energies of the original oscillators.


Our response to Point 1 is based on the following four observations.

\ben

\item The simple model discussed in \cite{jetzer}
is neither a valid description of the shunting resistor nor of the
Josephson junction. Jetzer and Straumann make computations for van
der Waals forces, whereas the measured spectra in the Josephson
junctions are a consequence of a completely different effect, the
ac Josephson effect \cite{tinkham}. Oversimplified {\it theoretical}
models may not shed any light on the origin of {\it measured}
noise spectra in Josephson junctions.

\item What is {\it measured} in the experiment of
Koch et al.  \cite{koch} is the spectrum of current fluctutions in
the resistive shunt, mixed down at the Josephson frequency. The
fact that the {\it experimental} data in \cite{koch} is so closely
fit by Equation (\ref{eq:power}) is an indication that at low
temperatures there is a significant correspondence between the
behaviour of this superconducting device and the prediction of the
corresponding theoretical treatment.

Jetzer and Straumann claim, on the basis of their simplified model
for van der Waals forces, that the linear term $\hbar \omega/2$ in
Equation (\ref{eq:power}) cannot be due to vacuum fluctuations.  It
may be a matter of semantics to argue about what to call the source
of this term, but their contention contradicts the received wisdom
\cite{gardiner,levinson,kogan} which clearly  singles out zero-point
fluctuations as the  source underlying the linear term $\hbar
\omega/2$ in the spectrum.

\item Arguments for  why  vacuum (zero-point) fluctuations have a measurable effect
in Josephson junctions have  been given by various authors, e.g.
\cite{levinson}.  Namely, a driven Josephson junction is a
non-equilibrium system, and non-equilibrium systems can be
influenced by vacuum fluctuations in a measurable way. For example,
zero-point fluctuations can cause excited atoms to return to the
ground state, thus producing an experimentally detectable  effect.
The argument against this observation presented in \cite{jetzer}
is based on equilibrium statistical mechanics and does not
incorporate non-equilibrium effects.

\item What is really at the root of the measured noise spectra in
resistors is the fluctuation dissipation theorem
\cite{kogan,callen,senitzky} which precisely predicts a power
spectrum as given by Equation (\ref{eq:power}). This spectrum has
been experimentally confirmed by Koch et al. \cite{koch} up to
frequencies of 0.6 THz.  All textbooks \cite{gardiner,kogan} and
classical papers \cite{callen,senitzky} on the subject emphasize the
fact that the linear term in the spectrum is induced by zero-point
fluctuations. \een Based on these points, we find Point 1 made by
Jetzer and Straumannn to be unconvincing.

\section{Point 2}\label{sec:pt2}
Turning to Point 2, it is clear that experiments involving van der
Waals forces or the Casimir effect can only probe differences in
vacuum energy. This is well known   and related to the fact that QED
is a renormalizable theory. Adding an arbitrary constant to the
vacuum energy density leaves the physical predictions of this theory
invariant. The correct conclusion is that experiments based
on the Casimir effect have no chance of measuring
the absolute value of vacuum energy.

The Josephson junction experiment, however, exploits a different
effect which apparently has nothing to do with the Casimir effect.
The theory of dissipative non-equilibrium quantum systems, such as
driven Josephson junctions, is much less well understood than the
Casimir effect. Whether the dissipative quantum theory underlying
resistively shunted Josephson junctions can be renormalized is
presently unclear. Hence the absolute value of vacuum energy may
well have physical meaning for these kinds of superconducting
quantum systems.

To illustrate this point,  assume that only differences in vacuum
energy are relevant for the Josephson junction experiment of Koch et
al., as Point 2 of Jetzer and Straumann suggests. It should  then be possible to add an
arbitrary constant (with the dimension of energy)  to Equation
(\ref{eq:energy}), without changing the physical predictions of the
theory. In our case the underlying theory is provided by the
fluctuation dissipation theorem \cite{callen,gardiner,kogan} which
predicts in complete generality that the mean square fluctuations
$\langle V^2 \rangle$ of the voltage in the shunting resistor are
given by
\begin{equation}
<V^2> = \frac{2}{\pi} \int \bar{U} (\omega/2\pi , T)
R(\omega) d\omega
\end{equation}
where $
\bar{U} (\nu , T)$ is  given by  Equation (\ref{eq:energy}) and
$R(\omega)$ is the shunting resistor. If we change $ \bar{U}$
by an additive constant $C$ to
\begin{equation}
\tilde{U} (\omega/2\pi,T) =\frac{1}{2}\hbar \omega +C + \frac{\hbar
\omega}{exp(\hbar \omega/kT)-1}, \label{uuu}
\end{equation}
the result would contradict the results of the Koch et al. \cite{koch}
experiment. 
Any $C \neq 0$ would imply voltage fluctuations in the resistor
different from those actually measured. Hence we obtain
a contradiction if we apply point 2 of Jetzer and Straumann to our system.

We thus conclude that Point 2 is unclear  from a theoretical point
of view, and further that the resolution of this question cannot be
decided on purely theoretical grounds. Rather, further experimental
investigation is necessary. In \cite{bm} we suggested an
experimental  check to see  whether a cutoff in the measurable
spectrum could be observed near the critical frequency $\nu_c=2\pi
\omega_c=1.7$ THz corresponding to dark energy density. If such a
cutoff is observed, it would indeed be the {\em new physics}
underlying this cutoff that makes the system couple to gravity and
make the absolute value of vacuum energy physically relevant.
Virtual photons that are not gravitationally active may well exist
beyond this cutoff, it is just the gravitationally active part of
vacuum fluctuations that would cease to exist at $\nu_c$.

A repeat of the Koch experiment, based on new types of Josephson
junctions operating in the THz region, will now be carried out by
Warburton \cite{warburton} and Barber and Blamire \cite{barber}.
These new experiments will measure the noise spectrum up to
frequencies exceeding the predicted critical frequency $\nu_c=1.7$
THz corresponding to the inferred dark energy density, using both
nitride and cuprate based Josephson junctions. This is an
interesting experimental project since the fluctuation dissipation
theorem and its potential contribution to dark energy density has
never been tested before at these high frequencies.

\section{Zeropoint energies and the fluctuation dissipation
theorem}

Jetzer and Straumann have recently published a new version of their
criticism \cite{jetzer2}. They now consider the fluctuation
dissipation theorem rather than an equilibrium van der Waals model,
thus adoping our point of view of what the relevant dynamics should
be. However, their main conclusion, printed in italics in their
concluding remarks, is still erroneous in our opinion. We quote

\begin{quotation}[Jetzer-Straumann, \cite{jetzer2}, p.58]
`Zero-point energies do not show up in any application of the
fluctuation dissipation theorem'.
\end{quotation}

Based on the above statement, Jetzer and Straumann again strongly
criticize our hypotheses.

Here we want to point out that the above statement of Jetzer and
Straumann, on which their entire criticism is based, is in sharp
contrast to the common interpretation taken  in the field. To
illustrate this point, let us provide a few quotations to show how
the universal term
\begin{equation}
H_{uni}:=\left[ \frac{1}{2}\hbar \omega + \frac{\hbar
\omega}{e^{\hbar \omega/kT}-1}\right] \label{huni}
\end{equation}
occurring in the fluctuation dissipation theorem is usually
interpreted physically (the emphasis in {\it italics} below is
added):

\begin{quotation}[Landau-Lifshitz \cite{landau}, p. 386]
`It should be noted that the factor in  the braces of
eq.~(\ref{huni}) is the {\em mean energy} (in units of $\hbar
\omega$) of an oscillator at temperature $T$; the term
$\frac{1}{2} \hbar \omega$ corresponds to the {\em zero-point
oscillations}.'
\end{quotation}

\begin{quotation}[Kogan \cite{kogan}, p. 55]
`Eq.~(\ref{huni}) describes the mean number of quanta (discrete
excitations) of an oscillator with frequency $\omega$ at
temperature $T$. The r.h.s. is the mean energy of this
oscillator. It consists of {\em ground state energy} $\hbar
\omega/2$ (it is called {\em zero-point energy}, or the energy of
zero-point vibrations) and the mean energy of the oscillator's
excitations.'
\end{quotation}

\begin{quotation}[Gardiner \cite{gardiner}, p. 5]
`The spectrum rises linearly with increasing $\omega$ because of
the first term in eq.~(\ref{huni}), which arises from the {\em
zeropoint fluctuations} in the {\em harmonic oscillators}...'
\end{quotation}

The main conclusion of Jetzer and Straumann in \cite{jetzer2},
quoted above, contradicts the standard view. All sources clearly
emphasize the fact that the linear term of the function $H_{uni}$
that occurs in the fluctuation dissipation theorem, connecting
fluctuation spectra with dissipation, has the physical meaning of a
zeropoint energy of a suitable quantum mechanical oscillator. We
thus think that the view of Jetzer-Straumann expressed in
\cite{jetzer2} is untenable.


On a closer inspection of \cite{jetzer2}, the reason why the authors
arrive at their non-standard view is immediately apparent. In
\cite{jetzer2}, the role of the universal function $H_{uni}$
occuring in the fluctuation dissipation theorem and the system
Hamiltonian $H_{sys}$ describing the quantum system under
consideration is confused. The authors re-derive in \cite{jetzer2}
the well-known fact that the fluctuation dissipation theorem is
valid for arbitrary Hamiltonians $H_{sys}$, in particular for those
where an arbitrary additive constant is added to $H_{sys}$. However,
their argument relates to the system Hamiltonian $H_{sys}$ and not
to $H_{uni}$. The idea that we proposed in \cite{bm} and further
worked out in \cite{bmnew} was to test in future experiments
\cite{warburton, barber} whether the zero-point term occuring in the
universal Hamiltonian $H_{uni}$ has any relation to dark energy. If
that is the case, a cutoff must be found in Josepshon experiments.
The zeropoint term in $H_{uni}$ cannot be removed by adding
arbitrary constants to it, as shown in section V. The line of
reasoning of Jetzer and Straumann in \cite{jetzer2} is
highly misleading in this context, since they add constants to a
different Hamiltonian, $H_{sys}$, which has nothing to do with the
universal Hamiltonian we consider, $H_{uni}$. In particular, the
considerations in \cite{jetzer2} provide no insight into the
physical interpretation of the vacuum energy associated with
$H_{uni}$, which is invariant and universal.

Jetzer and  Straumann state in \cite{jetzer2} that our insertion of
an arbitrary constant $C$  in eq.~(\ref{uuu}) is wrong. However,
they fail to explain to the reader that we did this 
for the sole purpose of {\em deriving a contradiction} of the
Jetzer-Straumann suggestion in \cite{jetzer}, namely to shift the
zeropoint energy of our system by adding an additive constant. So
certainly this equation is wrong, because it was our purpose to
derive a contradiction.

One remark is at order. Models of dark energy always require new
physics in one way or another. The question is where and in which
form this new physics enters. The class of models that can be
tested with Josephson junctions associate dark energy with
ordinary electromagnetic vacuum energy \cite{bmnew, review}.
Clearly, in order to reproduce the correct dark energy density in
the universe, for these types of models there must be a phase
transition point at around 1.7 THz where virtual photons loose
their gravitational activity. Virtual photons can still persist at
higher frequencies (hence ordinary QED is still valid), just
their {\em gravitational activity} ceases to exist at higher
frequencies in these types of models, 
by means of a phase transition describing a
change of gravitational behaviour of virtual photons at high
frequencies. This phase transition is the new physics
associated with the model. Since the zero-point term of $H_{uni}$
cannot be renormalized away, the above phase transition might be
observable in dissipative quantum systems described by the
fluctuation dissipation theorem. For this reason we think it is
very interesting to experimentally test the fluctuation
dissipation theorem at high frequencies, to either confirm or
rule out these types of dark energy models.


\section{Conclusions}

We have argued in this note that the objections  of Jetzer and
Straumann \cite{jetzer} to  the hypothesis formulated in \cite{bm}
are not applicable to our system. The arguments presented in
\cite{jetzer} are based on a model for the van der Waals forces
between two harmonic oscillators, which have nothing to do with the
measured noise spectra in Josephson junctions. Moreover, the
arguments presented in the more recent paper \cite{jetzer2} are in
apparent contrast to the standard textbook view on quantum noise in
Josephson junctions.

We further contend that the {\it only} way to really test the
hypothesis that there is a cutoff in the frequency spectrum of
measurable vacuum fluctuations is to actually do the experiment.
Appeal to theoretical arguments extended to situations in which the
theory has not been verified do not shed any light on the (so far
unknown) nature of dark energy.

\begin{acknowledgments}
This work was supported by the Engineering and Physical Sciences
Research Council (EPSRC, UK), the Natural Sciences and Engineering
Research Council (NSERC, Canada) and the Mathematics of
Information Technology and Complex Systems (MITACS, Canada). The
research was carried out while MCM was visiting the Mathematics
Institute of the University of Oxford.
\end{acknowledgments}


\begin{thebibliography}{99}
\bibitem{bm} C. Beck and M.C. Mackey, Phys. Lett. B {\bf 605}, 295
(2005) [astro-ph/0406504]
\bibitem{koch} R.H. Koch, D. van Harlingen, D. and
J. Clarke, Phys. Rev. B {\bf 26}, 74 (1982)
\bibitem{kochtheory} R.H. Koch, D. van Harlingen, and J. Clarke,
Phys. Rev. Lett. {\bf 45}, 2132 (1980)
\bibitem{callen} H. Callen and T. Welton, Phys. Rev. {\bf 83}, 34
(1951)
\bibitem{jetzer} P. Jetzer and N. Straumann, Phys. Lett. B {\bf 606},
77 (2005) [astro-ph/0411034]
\bibitem{jetzer2} P. Jetzer abd N. Straumann, Phys. Lett. B {\bf
639}, 57 (2006) [astro-ph/0604522]
\bibitem{landau} L.D. Landau and E.M. Lifshitz,
{\em Statistical Physics, Part 1, Landau Lifshitz Course of
Theoretical Physics, vol.5 }, Elsevier, Amsterdam (1980)
\bibitem{gardiner} C.W. Gardiner, {\em Quantum Noise},
Springer, Berlin (1991)
\bibitem{kogan} Sh. Kogan, {\em Electronic Noise and Fluctuations in
Solids}, Cambridge University Press, Cambridge (1996)
\bibitem{tinkham} M. Tinkham, {\em Introduction to
Superconductivity}, Dover Publications, New York (2004)
\bibitem{levinson} Y. Levinson, Phys. Rev. B {\bf 67},
184504 (2003)
\bibitem{senitzky} I. Senitzky, Phys. Rev. {\bf 119}, 670 (1960)
\bibitem{warburton} P.A. Warburton,
EPSRC (UK) grant EP/D029783/1 (2006)
\bibitem{barber} Z. Barber and M. Blamire,
EPSRC (UK) grant EP/D029872/1 (2006)
\bibitem{bmnew} C. Beck and M.C. Mackey, astro-ph/0605418, to appear in Physica A
\bibitem{review} E.J. Copeland, M. Sami, and S. Tsujikawa,
hep-th/0603057
\end{thebibliography}

\end{document}